\documentclass[aps,prd,showkeys,showpacs,twocolumn,a4paper]{revtex4-1}
\usepackage{ulem}
\usepackage{color}
\usepackage{graphicx}
\usepackage{hyperref}   
\usepackage{appendix}
\usepackage{epsfig}
\usepackage{epstopdf}
\usepackage{amsmath}
\usepackage{subfig}
\usepackage{comment}
\usepackage[dvipsnames]{xcolor}
\newcommand{\be}{\begin{equation}}
\newcommand{\ee}{\end{equation}}
\newcommand{\ba}{\begin{eqnarray}}
\newcommand{\ea}{\end{eqnarray}}

\begin{document}

\title{Thermal transport in a weakly magnetized hot QCD medium}
\author{Manu Kurian}
\email{manu.kurian@iitgn.ac.in}
\affiliation{Indian Institute of Technology Gandhinagar, Gandhinagar-382355, Gujarat, India}


\begin{abstract}

The thermal transport coefficients in a weakly magnetized quark-gluon plasma have been investigated within the ambit of a quasiparticle model to encode the effects of the realistic equation of  state.  The presence of a weak magnetic field leads to the Hall-type conductivity associated with thermal transport in the medium. An effective covariant kinetic theory has been employed to quantify the thermal dissipation while incorporating the mean field contributions in the medium. The interplay of thermal transport and electric charge transport in the weakly magnetized medium has been explored in terms of the Wiedemann-Franz law. Strong violation of the Wiedemann-Franz law has been observed in temperature regimes near to the transition temperature. The behaviour of thermal conductivity in the strong magnetic field limit has also been studied. It is observed that both the magnetic field and equation of state have a significant impact on the thermal dissipation in the medium.

\end{abstract}


\keywords{Thermal conductivity, Quark-gluon plasma, Weak magnetic field, Effective fugacity, Electrical conductivity, Wiedemann-Franz law}

\maketitle

 \section{Introduction}
 
It is expected that an intense magnetic field has been generated in the very initial stages of non-central collisions at Relativistic Heavy-Ion Collider (RHIC) and the Large Hadron Collider (LHC)~\cite{Skokov:2009qp,Zhong:2014cda,Voronyuk:PRC2011,Deng:PRC2012}. 
The measurement of the RHIC~\cite{Adam:2019wnk} and the very recent LHC observation~\cite{Acharya:2019ijj} on the directed flow $v_1$ for charged hadrons and $D/\bar{D}^0$ mesons confirms the existence of the strong magnetic fields in the high energetic collisions. However, a clear picture of the evolution of the magnetic field and its lifetime in the medium is yet to be known. Studies have shown that the magnetic field decays rapidly in a vacuum, and the evolution of the field in the system of charged particles may depend on the medium properties, say, electrical conductivity~\cite{Tuchin:PRC882013,McLerran:NPA9292014}. This suggests that the magnetic field may sustain in the hot QCD medium for a longer time than anticipated. 

The magnetic field may affect the transport and thermodynamical behaviour of the quark-gluon plasma (QGP), created in the heavy-ion collisions~\cite{Koothottil:2018akg,Dey:2019vkn,Astrakhantsev:2019zkr}.
In particular, anomalous transport phenomena~\cite{Fukushima:2008xe,Sadofyev:2010pr,She:2017icp,Kharzeev:2015znc}, magnetic catalysis~\cite{Gusynin:1995nb}, electromagnetic probes~\cite{Bandyopadhyay:PRD2016,Tuchin:PRC832011}, quarkonia suppression~\cite{Hasan:EPJC2017,Singh:PRD972018}, heavy quark transport~\cite{Fukushima:2015wck,Singh:2020faa}, and jet quenching~\cite{Li:2016bbh} in the magnetized medium have gained huge momentum in recent years. There have been several investigations of the QGP medium properties in the presence of a strong magnetic field~\cite{Karmakar:2019tdp,Hattori:2017qih,Kurian:2018qwb,Rath:2019vvi,Fukushima:2017lvb} and also in a weak magnetic field limit~\cite{Feng:PRD962017,Ghosh:2018cxb,Bandyopadhyay,Das:2019ppb}. Notably, in a strongly magnetized medium, the charged fermions follow $1+1-$dimensional Landau level dynamics along the direction of the magnetic field. Whereas in a weakly magnetized medium, the temperature is the dominant energy scale of the system, and the magnetic field effects enter through the cyclotron frequency of the charged particles in the medium. 

The thermal dissipation in the medium is due to temperature gradient over the spatial separations of fluid and can be described in terms of the transport coefficient, thermal conductivity, for a system with conserved baryon current density. In Refs.~\cite{Denicol:2012vq,Kapusta:2012zb}, the authors have analyzed the significance of thermal conductivity in the relativistic dissipative hydrodynamical expansion of the medium. Thermal conductivity has been studied within Kubo formalism~\cite{FernandezFraile:2009mi}, Nambu-Jona-Lasinio (NJL) Model~\cite{Marty:2013ita,Deb:2016myz}, transport model~\cite{Greif:2013bb}, and kinetic theory approach~\cite{Mitra:2017sjo,Kalikotay:2019fle}. The electrical and thermal conductivities for the hadronic medium within the scope of a hadron resonance gas model have also been estimated~\cite{Kadam:2017iaz}. The thermoelectric behavior of the hot nuclear matter and the associated Seebeck coefficient have started receiving much attention very recently~\cite{F,Dey:2020sbm,Das:2020beh,Zhang:2020efz}. The authors of Refs.~\cite{Feng:PRD962017,Das:2019ppb,Das:2019wjg,Das:2019pqd} have studied the electrical conductivity and Hall conductivity in a weakly magnetized medium at finite quark chemical potential. It is important to emphasize that the Hall current, that is transverse to the electric and magnetic field, vanishes in the strong magnetic field limit due to the $1+1-$dimensional Landau kinematics of the charged particles. The longitudinal heat current and the associated thermal conductivity of the QGP in the presence of a strong magnetic field have been investigated in Refs.~\cite{Kurian:2018qwb,Rath:2019vvi}. It is an interesting task to extend the analysis of heat current to a weakly magnetized QGP medium to study all components of heat current in the medium. 

The current work primarily focuses on the effective description of the thermal transport of a weakly magnetized QGP medium. The hot QCD medium interactions are incorporated in the analysis through the effective modeling of quarks/antiquarks and gluonic degrees of freedom employing an effective fugacity quasiparticle model (EQPM)~\cite{Chandra:2011en,Chandra:2007ca,Kurian:2017yxj}. The EQPM is successful in describing the medium transport coefficients~\cite{Mitra:2017sjo}, heavy quark dynamics~\cite{Das:2012ck} and dilepton production~\cite{Chandra:2015rdz} in the QGP medium.
The description of thermal dissipation requires the knowledge of the system away from thermal equilibrium. To that end, the effective Boltzmann equation has been solved within the relaxation time approximation~\cite{Mitra:2018akk}. As the magnetic field is the subdominant energy scale compared to the temperature scale, the magnetic field effects are entering through the Lorenz force term in the Boltzmann equation. The relative behaviour of thermal transport and electrical transport has been investigated by employing Wiedemann-Franz law for the weakly magnetized medium.     

The paper is organized as follows. Section II describes the formalism of thermal transport in the QGP in the presence of a weak magnetic field within a quasiparticle description. Section III is devoted to the relative significance of the thermal transport and electric charge transport in a weakly magnetized medium, followed by the comparison of thermal conductivity in the weak and strong magnetic field regimes.   Results and discussions of the analysis are presented in section IV. Finally, we summarize with an outlook in section V.\

{\bf Notations and conventions:} In the current analysis, we define the metric tensor as $g^{\mu\nu}=$diag$(1 ,-1, -1, -1)$. The fluid velocity $u^{\mu} = (1, 0, 0, 0)$ is normalized to unity in the rest frame.  The fractional charge of the quark is $q_f=2e/3, -e/3, -e/3$ for up, down, and strange quarks, respectively. We define the projection operator as $\Delta^{\mu\nu}\equiv g^{\mu\nu}-u^\mu u^\nu$ and is orthogonal to $u^{\mu}$. The four-index traceless symmetric projection operator takes the form, $\Delta^{\mu\nu}_{\alpha\beta}\equiv\frac{1}{2}(\Delta^\mu_\alpha\Delta^\nu_\beta +\Delta^\mu_\beta\Delta^\nu_\alpha)-\frac{1}{3}\Delta^{\mu\nu}\Delta_{\alpha\beta}$. The index $k$ denotes the particle species and $g_k$ is the degeneracy factor in the present analysis. For quarks and antiquarks, $g_{q,\bar{q}}=\sum_f2N_c$ such that $g_{q,\bar{q}}=2N_cN_f$ in the absence of magnetic field, where $N_f$ is the number of flavor.

\section{Thermal conductivity in the presence of a weak magnetic field}

\subsubsection*{EQPM description of thermal conductivity with finite mass and quark chemical potential at $B=0$}

The EQPM equilibrium distribution function of quarks/antiquarks and gluons, with a small but finite quark chemical potential $\mu$ take the following forms~\cite{Bhadury:2019xdf},
\begin{align}\label{1}
f^0_{q/\bar{q}} &=\frac{z_q \exp{[-\beta (u\!\cdot\! p_q\mp \mu)]}}{1 + z_q\exp{[-\beta (u\!\cdot\! p_q \mp \mu)]}},\\
f^0_g &=\frac{z_g \exp{[-\beta\, u\!\cdot\! p_g]}}{1 - z_g\exp{[-\beta\, u\!\cdot\! p_g]}},
\end{align}
where $z_q$ and $z_g$ are the temperature dependent effective fugacity parameter of the quarks/antiquarks and gluons, respectively. These parameters encode the thermal medium effects via the $(2+1)-$flavor lattice EoS in the effective description of the QGP medium.
Note that the effective fugacities are not connected with any conserved number current in the medium, and the temperature dependence of the fugacity parameter remains the same with the finite but small $\mu-$limit~\cite{Mitra:2017sjo}. Hence, the fugacity parameter for quark and antiquark is same, $i.e.$, $z_q=z_{\bar{q}}$~\cite{Chandra:2011en, Chandra:2007ca}. Effective fugacities relates the dressed particle (quasiparticle) four-momenta $\Tilde{p}_k^{\mu}$ and bare particle four-momenta $p_k^{\mu}$ as,
\begin{equation}\label{2}
\Tilde{p_k}^{\mu} = p_k^{\mu}+\delta\omega_k\, u^{\mu}, \qquad
\delta\omega_k= T^{2}\,\partial_{T} \ln(z_{k}),
\end{equation}
which defines the quasiparticle energy $\omega_{k}$ as,
\begin{equation}\label{3}
\Tilde{p_k}^{0}\equiv\omega_{k}=\epsilon_k+\delta\omega_k,
\end{equation}
where the bare particle energy $\epsilon_k={\sqrt{\mid{\bf{\Tilde{p}}}_k\mid^2+m_q^2}}$ for quarks/antiquarks and $\epsilon_k=\mid{\bf{\Tilde{p}}}_k\mid$ for gluons.
The EQPM definition of the energy momentum tensor $T^{\mu\nu}$ and four current $N^{\mu}$ in terms of dressed momenta ${\bf{\Tilde{p}}}_k$ take the following forms~\cite{Mitra:2018akk},
\begin{align}\label{4}
T^{\mu\nu}(x)&=\sum_{k}g_k\int{d\Tilde{P}_k\,\Tilde{p}_k^{\mu}\,\Tilde{p}_k^{\nu}\,f_k(x,\Tilde{p}_{k})}\nonumber\\
&+\sum_{k}\delta\omega_k\,g_k\int{d\Tilde{P}_k\,\frac{\langle\Tilde{p}_k^{\mu}\,\Tilde{p}_k^{\nu}\rangle}{\epsilon_k}\, f_k(x,\Tilde{p}_{k})},
\end{align}
and
\begin{align}\label{5}
N^{\mu}(x)&=\sum_{k}g_k\int{d\Tilde{P}_k\,\Tilde{p}_k^{\mu}\,f_k(x,\Tilde{p}_{k})}\nonumber\\
&+\sum_{k}\delta\omega_k\,g_k\int{d\Tilde{P}_k\,\frac{\langle\Tilde{p}_k^{\mu}\rangle}{\epsilon_k}\, f_k(x,\Tilde{p}_{k})},
\end{align}
respectively, with the integral measure $d\Tilde{P}_k\equiv\frac{d^3\mid{\bf{\Tilde{p}}}_k\mid}{(2\pi)^3\omega_{k}}$. Here, $\langle\Tilde{p}_k^{\mu}\,\Tilde{p}_k^{\nu}\rangle\equiv\frac{1}{2}(\Delta^{\mu}_{\alpha}\Delta^{\nu}_{\beta}+\Delta^{\mu}_{\beta}\Delta^{\nu}_{\alpha})\,\tilde{p}_k^{\alpha}\,\Tilde{p}_k^{\beta}$  and $\langle\Tilde{p}^{\mu}_q\rangle\equiv\Delta^\mu_\nu\,\Tilde{p}^{\nu}_q$ is the rank one irreducible tensor.

For the system near to local thermodynamic equilibrium, the quasiparticle momentum distribution function takes the form $f_k=f^0_k+\delta f_k$, where $\delta f_k/f^0_k \ll1$ for the $k$-th species. Thus the macroscopic quantities can be defined in terms of equilibrium and non-equilibrium parts, $T^{\mu\nu}=T^{0~\mu\nu}+\Delta T^{\mu\nu} (x)$ and $N^{\mu}=N^{0~\mu}+\Delta N^{\mu} (x)$ with,
\begin{align}\label{6}
\Delta T^{\mu\nu}(x)&=\sum_{k}g_k\int{d\Tilde{P}_k\,\Tilde{p}_k^{\mu}\,\Tilde{p}_k^{\nu}\,\delta f_k(x,\Tilde{p}_{k})}\nonumber\\
&+\sum_{k}\delta\omega_k\,g_k\int{d\Tilde{P}_k\,\frac{\langle\Tilde{p}_k^{\mu}\,\Tilde{p}_k^{\nu}\rangle}{\epsilon_k}\, \delta f_k(x,\Tilde{p}_{k})},
\end{align}
and
\begin{align}\label{7}
\Delta N^{\mu}(x)&=\sum_{k}g_k\int{d\Tilde{P}_k\,\Tilde{p}_k^{\mu}\,\delta f_k(x,\Tilde{p}_{k})}\nonumber\\
&+\sum_{k}\delta\omega_k\,g_k\int{d\Tilde{P}_k\,\frac{\langle\Tilde{p}_k^{\mu}\rangle}{\epsilon_k}\, \delta f_k(x,\Tilde{p}_{k})}.
\end{align}
The covariant effective Boltzmann equation describes the evolution of quasiparticle distribution function and has the following form, 
\begin{equation}\label{8}
\Tilde{p}^{\mu}_k\,\partial_{\mu}f_k(x,\Tilde{p}_k)+F_k^{\mu}\left(u\!\cdot\!\tilde{p}_k\right)\partial^{(p)}_{\mu} f_k = C[f_{k}]\equiv -\left(u\!\cdot\!\tilde{p}_k\right)\frac{\delta f_k}{\tau_{R_k}},
\end{equation}
where $C[f_k]$ is the collision integral and is defined in terms of thermal relaxation time $\tau_{R}$ within the relaxation time approximation (RTA)~\cite{Anderson_Witting}. The quantity $F_k^{\mu}=-\partial_{\nu}(\delta\omega_k u^{\nu}u^{\mu})$ is the mean field force term that defined from the conservation of the EQPM energy momentum and particle four-flow.
We solve the relativistic transport equation within RTA to obtain $\delta f_k$ by taking an iterative Chapman-Enskog like solution~\cite{Jaiswal:2013npa} for multicomponent, many particle system. The first order correction the distribution function takes the forms, 
\begin{align}\label{9}
\!\!\delta f_k &= \tau_{R_k}\Bigg[\frac{1}{T} \bigg\{\Tilde{p}_k^0~\partial_0 T + \Tilde{p}_k^i\,\partial_i T \bigg\} \!+\! \frac{T}{\Tilde{p}^0_k} \!\bigg\{\Tilde{p}^0_k\,\partial_0\Big(\frac{\mu}{T}\Big) \nonumber\\
&\!+\! \Tilde{p}_k^i\,\partial_i \Big(\frac{\mu}{T}\Big) \bigg\}\!-\! \frac{1}{\Tilde{p}^0_{k}} \!\bigg\{\Tilde{p}^0_{k}\, \Tilde{p}_{k}^\nu\,\partial_0 u_\nu \!+\! \Tilde{p}^i_{k}\, \Tilde{p}_{k}^\nu\,\partial_i u_\nu \bigg\}\!+\! \theta\,\delta\omega_k \Bigg]\frac{\partial f^0_k}{\partial\epsilon_k}.
\end{align}
The non-equilibrium part of the distribution function encodes different thermodynamic forces that correspond to different transport processes. The trace part of velocity gradient $\theta=\partial_\mu u^\mu$ and traceless part of velocity gradient $\Delta^{\mu\nu}_{\alpha\beta}\partial_{\alpha}u_{\beta}$ denotes the   bulk and shear viscous force, respectively. The current focus is on the thermal driving force, which is related to the temperature gradient in the medium. It is important to emphasize that these thermodynamic forces are independent to each other. Employing $\partial_0 u_\nu=\frac{\nabla_\nu P}{nh}$ from the energy-momentum conservation, where $n$ is the number density and $h=\frac{\varepsilon+P}{n}$ is the enthalpy per particle, along with the relativistic Gibbs-Duhem relation, $\partial_i \Big(\frac{\mu}{T}\Big)=-\frac{h}{T^2}(\partial_iT-\frac{T}{nh}\partial_iP)$ for the system of multicomponent many particles, we have

\begin{align}\label{10}
\!\!\delta f_k = \tau_{R_k}\frac{\partial f^0_k}{\partial\epsilon_k}\big(\omega_{k}-h_k\big){\bf v}_k.{\bf X} + \delta f_{k\,{\text{shear}}}+\delta f_{k\,{\text{bulk}}},
\end{align}
where $v^i=\frac{\Tilde{p}^i}{\Tilde{p}^0}$ and the thermal driving force takes the form as follows,
\begin{align}\label{11}
X_{i}=\frac{\partial_i T}{T}-\frac{\partial_i P}{nh}.
\end{align}
Note that in the steady state, the momentum conservation implies that $\partial_i P=0$. Thermal conduction involves the relative flow of energy, and the heat current for single component particle $I_k^i$ takes the form, 
\begin{align}\label{12}
I_k^{i}=\Delta T_k^{0i}-h_k\Delta N_k^{i},
\end{align}
where the $\Delta T^{0i}$ and $\Delta N^{i}$ take the forms as follows, 
\begin{align}\label{13}
\Delta T^{0i}&=\sum_{k}g_k\frac{\tau_{R_k}}{3}\int{d\Tilde{P}_k\,\mid{\bf{\Tilde{p}}}_k\mid^2 \, \big({\omega_{k}-h_k}\big)\frac{\partial f^0_k}{\partial\epsilon_k}X^i},
\end{align}
and
\begin{align}\label{14}
\Delta N^{i}&=\sum_{k}g_k\frac{\tau_{R_k}}{3}\int{d\Tilde{P}_k\,\frac{\mid{\bf{\Tilde{p}}}_k\mid^2}{\omega_k} \, \big({\omega_{k}-h_k}\big)\frac{\partial f^0_k}{\partial\epsilon_k}X^i}\nonumber\\
&-\sum_{k}\delta\omega_k g_k\frac{\tau_{R_k}}{3}\int{d\Tilde{P}_k\,\frac{\mid{\bf{\Tilde{p}}}_k\mid^2}{\omega_k~\epsilon_k} \, \big({\omega_{k}-h_k}\big)\frac{\partial f^0_k}{\partial\epsilon_k}X^i}.
\end{align}
One can define the heat conductivity $\kappa$ employing either the Eckart or Landau-Lifshitz condition as, $I^i=-\kappa TX^{i}$. Employing the Gibbs-Duhem relation, one can rewrite the definition of heat current as, $I^i=\kappa \frac{T^2}{h}\partial_{i}(\frac{\mu}{T})$. 
This implies that the thermal conductivity vanishes for a system without any conserved current. Substituting Eq.~(\ref{13}) and Eq.~(\ref{14}) on the definition of heat current, we obtain thermal conductivity as, 
\begin{align}\label{15}
\kappa&=\frac{1}{3T}\sum_{k}g_k{\tau_{R_k}}\int{d\Tilde{P}_k\,\frac{\mid{\bf{\Tilde{p}}}_k\mid^2}{\omega_k} \, \big({\omega_{k}-h_k}\big)^2\Big(-\frac{\partial f^0_k}{\partial\epsilon_k}\Big)}\nonumber\\
&+\frac{1}{3T}\sum_{k}\delta\omega_kg_k{\tau_{R_k}}\int{d\Tilde{P}_k\,\frac{\mid{\bf{\Tilde{p}}}_k\mid^2}{\omega_k~\epsilon_k} \, h_k\big({\omega_{k}-h_k}\big)\Big(-\frac{\partial f^0_k}{\partial\epsilon_k}\Big)}.
\end{align}
Note that in the ultra-relativistic limit ($z_k=1$), the expression of thermal conductivity reduces back to that in Ref.~\cite{Chakrabarty:1986xx}. The Eq.~(\ref{15}) denotes the EQPM description of thermal conductivity in the absence of the magnetic field in the medium. 

\subsubsection*{Thermal conductivity in a weak magnetic field}

As the strength of the magnetic field is weak, particle dispersion is not directly affected by the field, unlike the $1+1-$dimensional Landau level kinematics in the strong field limit. The effective relativistic Boltzmann equation in the presence of electromagnetic field strength tensor $F^{\mu\nu}$ modifies to the following form, 
\begin{align}\label{16}
\Tilde{p}^{\mu}_k\,\partial_{\mu}f_k(x,\Tilde{p}_k)+\bigg(F_k^{\mu}\left(u\!\cdot\!\tilde{p}_k\right)+&q_{f_k}F^{\mu\nu}\Tilde{p}_{k\,  \mu}\bigg)\partial^{(p)}_{\mu} f_k\nonumber\\ &=-\left(u\!\cdot\!\tilde{p}_k\right)\frac{\delta f_k}{\tau_{R_k}},
\end{align}
where $q_{f_k}$ is the charge of the particle of flavor $f$. It has been observed in the previous studies~\cite{Kurian:2018qwb,Kurian:2017yxj} that the magnetic field has a strong dependence on QCD thermodynamics and collision kernel in the strong field limit. This is attributed to the fact that the magnetic field is considered as the dominant energy scale in the system. In the current analysis, the weak magnetic field  can be considered as a small perturbation in the system. This allows us to ignore the effect of the magnetic field on the thermodynamics and collision integral in the present analysis. The EQPM description of thermal relaxation time for the $p_k+p_l\longrightarrow p_{k^{'}}+p_{l^{'}}$ binary scattering processes is described in detail in Ref.~\cite{Mitra:2017sjo}. It turned out that the thermal relaxation depends on the coupling constant as,
\begin{align}\label{17}
    \tau^{-1}_{R\,{g,q,\bar{q}}}\sim T\alpha^2_{\text{eff}}\ln\bigg\{{\frac{1}{\alpha_{\text{eff}}}\bigg\}}.
\end{align}
The EQPM is based on charge renormalization, and effective coupling $\alpha_{\text{eff}}$ can be defined from the Debye screening of the hot QCD medium. The effective coupling at finite temperature and chemical potential takes the following form,
\begin{align}\label{18}
{\alpha_{\text{eff}}}=&
\Bigg[ \dfrac{2N_c}{\pi^{2}}
PolyLog[2,z_{g}]-\dfrac{2N_f}
{\pi^{2}}PolyLog[2,-z_{q}]
\nonumber\\
&+\mu^2 \frac{N_f}{\pi^2}\frac{z_q}{1+z_q}\Bigg] \dfrac{\alpha_{s}(T, \mu)}{\Big( \frac{N_c}{3}+\frac{N_f}{6}+\mu^2\frac{N_f}{2\pi^2}\Big)},
\end{align}
where $\alpha_{s}(T, \mu)$ is the running coupling constant taken from 2-loop QCD gauge coupling constants. Note that the same analysis will not hold true for the case of the strongly magnetized medium. The estimation of the collision kernel and relaxation time in a strongly magnetized medium has been investigated in Refs.~\cite{Kurian:2018qwb}. 

Since the current focus is only on the effects of thermal transport in the presence of a magnetic field and, taking account of the fact that all thermodynamic forces are independent, we omit all other forces other than thermal driving forces, as described in Refs.~\cite{Das:2020beh}. Expanding each term in the Boltzmann equation and keeping terms corresponding to the thermal transport in the weakly magnetized medium, we obtain 
\begin{align}\label{19}
\Big[-\frac{\partial f^0_k}{\partial\epsilon_k}\big({\omega_{k}-h_k}\big)\Big]{\bf v}_k.{\bf X} + q_{f_k} ({\bf v}_k\times {\bf B}).\frac{\partial f_k}{\partial{\bf{\Tilde{p}}}_k}=-\frac{\delta f_k}{\tau_{R_k}}.
\end{align}
%
The term $\frac{\partial f^0_k}{\partial{\bf{\Tilde{p}}}_k}\propto {\bf v}_k$ and hence the Lorenz force vanishes in the equilibrium case. Hence, we have 
\begin{align}\label{20}
\Big[-\frac{\partial f^0_k}{\partial\epsilon_k}\big({\omega_{k}-h_k}\big)\Big]{\bf v}_k.{\bf X} + q_{f_k} ({\bf v}_k\times {\bf B}).\frac{\partial \delta f_k}{\partial{\bf{\Tilde{p}}}_k}=-\frac{\delta f_k}{\tau_{R_k}}.
\end{align}
The Eq.~(\ref{20}) can be solved by choosing the following ansatz for the non-equilibrium part of the distribution function,
\begin{align}\label{21}
\delta f_k=({\bf{\Tilde{p}}}_k\, .\, {\bf \Xi})\,\frac{\partial f^0_k}{\partial\epsilon_k},
\end{align}
in which ${\bf \Xi}$ related to the thermal driving force and the magnetic field in the medium and takes the following form,
\begin{align}\label{22}
{\bf \Xi}=\alpha_1\, {\bf b}+\alpha_2\, {\bf X}+\alpha_3\, \Big({\bf X}\times{\bf b}\Big).
\end{align}
Here, ${\bf b}=\frac{{\bf B}}{\mid{\bf B}\mid}$ is the direction of the magnetic field in the medium. 
In Ref.~\cite{Dash:2020vxk}, the authors have estimated five components of shear stress tensor and two components of bulk viscosity in the magnetized medium.
Note that there will be more terms in the analysis if we switch on the electric field and viscous effects, which corresponds to seven viscous coefficients (five shear viscosity and two bulk viscosity) and two electric charge transport coefficients (electrical and hall conductivity) in the medium.  Substituting Eq.~(\ref{21}) in Eq.~(\ref{20}) and employing the general expression of ${\bf \Xi}$ as described in Eq.~(\ref{22}), we obtain,
\begin{widetext}
\begin{align}\label{23}
&-\big({\omega_{k}-h_k}\big){\bf v}_k.{\bf X}+q_{f_k} {\bf v}_k. ({\bf B}\times \alpha_2{\bf X})+q_{f_k} {\bf v}_k. \Big({\bf B}\times \alpha_3({\bf X}\times {\bf b} )\Big)=-\frac{\omega_k}{\tau_{R_k}}\bigg[\alpha_1{\bf v}_k.{\bf b}+\alpha_2{\bf v}_k.{\bf X}+\alpha_3{\bf v}_k.({\bf X}\times{\bf b})\bigg].
\end{align}
\end{widetext}
The parameters $\alpha_1$, $\alpha_2$, and $\alpha_3$ can be obtained by comparing the independent terms with different tensor structures in both sides of the Eq.~(\ref{23}), and we have,
\begin{align}\label{24}
& \frac{\omega_k}{\tau_{R_k}}\alpha_1=\alpha_3q_{f_k} \mid{\bf B}\mid ({\bf b}.{\bf X}),\\
& \frac{\omega_k}{\tau_{R_k}}\alpha_2=\big({\omega_{k}-h_k}\big)-\alpha_3q_{f_k} \mid{\bf B}\mid,\label{24.1}\\
& \frac{\omega_k}{\tau_{R_k}}\alpha_3=\alpha_2q_{f_k} \mid{\bf B}\mid.\label{24.2}
\end{align}
Employing Eqs.~(\ref{24}-\ref{24.2}) and defining $\Omega_{c\, k}=\frac{q_{f_k} \mid{\bf B}\mid}{\omega_k}$, where $\mid\Omega_{c\, k}\mid$ describes the cyclotron frequency, the parameters reduced to the following forms, 
\begin{align}\label{25}
&\alpha_1=\frac{\tau^2_R}{\omega_k}\frac{\big({\omega_{k}-h_k}\big)}{(1+\tau_{R_k}^2\, \Omega^2_{c\, k})}\, \Omega^2_{c\, k} \,({\bf b}.{\bf X}),\\
&\alpha_2=\frac{\tau_{R_k}}{\omega_k}\frac{\big({\omega_{k}-h_k}\big)}{(1+\tau_{R_k}^2\, \Omega^2_{c\, k})},\label{25.1}\\
&\alpha_3=\frac{\tau_{R_k}^2}{\omega_k}\frac{\big({\omega_{k}-h_k}\big)}{(1+\tau_{R_k}^2\, \Omega^2_{c\, k})}\,  \Omega_{c\, k}.\label{25.2}
\end{align}
Substituting Eqs.~(\ref{25}-\ref{25.2}) in Eq.~(\ref{22}), we obtain the non-equilibrium correction to the distribution function in the presence of the weak magnetic field from Eq.~(\ref{21}) as,
\begin{align}\label{26}
\delta f_k=&\tau_{R_k}\frac{\big({\omega_{k}-h_k}\big)}{(1+\tau_{R_k}^2\, \Omega^2_{c\, k})}\bigg[\big({\bf v}_k.{\bf X}\big)+\tau_{R_k}\, \Omega_{c\, k} {\bf v}_k. \big({\bf X}\times {\bf b}\big)\nonumber\\
&+\tau_{R_k}\, \Omega^2_{c\, k}\,\big({\bf b}.{\bf X}\big)\,\big({\bf v}_k.{\bf b}\big)\bigg]\,\frac{\partial f^0_k}{\partial\epsilon_k}.
\end{align}
Using Eq.~(\ref{26}), the heat current in the weakly magnetized medium takes the form as follows,
\begin{align}\label{27}
I^{i}&=\sum_{k}g_k{\tau_{R_k}}\int{d\Tilde{P}_k}\, {\Tilde{p}^i_k} \frac{\big({\omega_{k}-h_k}\big)^2}{(1+\tau_{R_k}^2\, \Omega^2_{c\, k})}\bigg[\big({\bf v}_k.{\bf X}\big)\nonumber\\
&+\tau_{R_k}\, \Omega_{c\, k} {\bf v}_k. \big({\bf X}\times {\bf b}\big)+\tau_{R_k}\, \Omega^2_{c\, k}\,\big({\bf b}.{\bf X}\big)\,\big({\bf v}_k.{\bf b}\big)\bigg]\,\frac{\partial f^0_k}{\partial\epsilon_k}\nonumber\\
&+\sum_{k}\delta\omega_kg_k{\tau_{R_k}}\int{d\Tilde{P}_k}\, \frac{\Tilde{p}^i_k}{\epsilon_k} \frac{h_k\big({\omega_{k}-h_k}\big)}{(1+\tau_{R_k}^2\, \Omega^2_{c\, k})}\bigg[\big({\bf v}_k.{\bf X}\big)\nonumber\\
&+\tau_{R_k}\, \Omega_{c\, k} {\bf v}_k. \big({\bf X}\times {\bf b}\big)+\tau_{R_k}\, \Omega^2_{c\, k}\,\big({\bf b}.{\bf X}\big)\,\big({\bf v}_k.{\bf b}\big)\bigg]\,\frac{\partial f^0_k}{\partial\epsilon_k}.
\end{align}
To quantify the effects of the magnetic field in thermal transport, we further simplified the analysis by fixing the direction of the magnetic field along the $z-$axis and the temperature gradient (thermal driving force) in the $x-y$ plane. Following this condition, the heat current takes the form as follows,
\begin{align}\label{28}
    {\bf I}=-\kappa_0 T {\bf X}-\kappa_1 T({\bf X}\times {\bf b}),
\end{align}
where the thermal transport coefficients in the weakly magnetized medium, $\kappa_0$ and $\kappa_1$, can be defined as,
\begin{align}\label{29}
\kappa_0=&\frac{1}{3T}\sum_{k}g_k\,{\tau_{R_k}}\int{d\Tilde{P}_k\,\frac{\mid{\bf{\Tilde{p}}}_k\mid^2}{\omega_k} \, \frac{\big({\omega_{k}-h_k}\big)^2}{(1+\tau_{R_k}^2\, \Omega^2_{c\, k})}\Big(-\frac{\partial f^0_k}{\partial\epsilon_k}\Big)}\nonumber\\
&+\frac{1}{3T}\sum_{k}\delta\omega_k\,g_k\,{\tau_{R_k}}\int{d\Tilde{P}_k}\,\frac{\mid{\bf{\Tilde{p}}}_k\mid^2}{\omega_k~\epsilon_k} \, \frac{h_k\big({\omega_{k}-h_k}\big)}{(1+\tau_{R_k}^2\, \Omega^2_{c\, k})}\nonumber\\
&\times\Big(-\frac{\partial f^0_k}{\partial\epsilon_k}\Big),
\end{align}
and 
\begin{align}\label{30}
\kappa_1=&\frac{1}{3T}\sum_{k}g_k\,{\tau_{R_k}^2}\int{d\Tilde{P}_k}\,\frac{\mid{\bf{\Tilde{p}}}_k\mid^2}{\omega_k} \, \frac{\big({\omega_{k}-h_k}\big)^2}{(1+\tau_{R_k}^2\, \Omega^2_{c\, k})}\, \Omega_{c\, k}\nonumber\\
&\times \Big(-\frac{\partial f^0_k}{\partial\epsilon_k}\Big)\nonumber\\
&+\frac{1}{3T}\sum_{k}\delta\omega_k\,g_k\,{\tau_{R_k}^2}\int{d\Tilde{P}_k}\,\frac{\mid{\bf{\Tilde{p}}}_k\mid^2}{\omega_k~\epsilon_k} \, \frac{h_k\big({\omega_{k}-h_k}\big)}{(1+\tau_{R_k}^2\, \Omega^2_{c\, k})}\nonumber\\
&\times\, \Omega_{c\, k}\,\Big(-\frac{\partial f^0_k}{\partial\epsilon_k}\Big),
\end{align}
respectively. At ${\bf{B}}=0$, Eq.~(\ref{29}) reduces back to the definition of conductivity in the absence of a magnetic field as in Eq.~(\ref{15}). The expression of heat current Eq.~(\ref{28}) is analogous to that of electric current in the magnetized medium and is discussed in detail in the next section.

\section{Relative significance of thermal transport in the medium}

This section deals with the relative importance of heat transport and electric charge transport in a weakly magnetized hot QCD medium. This can be quantified in terms of Lorenz number of the medium.  As the dynamics of particles are different in the presence of the strong magnetic field compared to the weakly magnetized system, we compare the thermal dissipative process and associated transport coefficients in both these regimes of the magnetic field. 
\subsection{Thermal transport versus charge transport in a weak magnetic field}

The relative importance of thermal and charge transport in the QCD medium can be understood in terms of Wiedemann-Franz law:
\begin{equation}\label{31}
\frac{\text{Thermal conductivity}}{T\times\text{Charge conductivity}}\equiv L,
\end{equation}
where $L$ is the Lorenz number. The magnitude of the $L$ describes the system as good electrical as well as thermal conductor. In general for anisotropic cases, the electrical conductivity and thermal conductivity are tensorial quantities, and hence $L$ has different components ($i.e.$, $L$ posses tensorial form). 
The main focus here is to study whether the QCD medium follows this behaviour in the presence of the weak magnetic field while incorporating QCD EoS effects via the EQPM. This requires the EQPM description of charge transport coefficients in the weakly magnetized medium. 

In the present analysis, we consider the electric field ${\bf{E}}$ in the direction transverse to that of the magnetic field, say along $x-$axis. The electric current ${\bf j}$ in the weakly magnetized medium can be defined as,
\begin{align}\label{32}
    {\bf j}=\sigma_e {\bf E} + \sigma_H  ({\bf E}\times {\bf b}),
\end{align}
where $\sigma_e$ and $\sigma_H$ denotes the electrical conductivity and Hall conductivity, respectively. The EQPM description of electric current while incorporating mean-field terms takes the following form,
\begin{align}\label{33}
{\bf {j}}&=\sum_f2N_cq_{f_ q}\int{d\Tilde{P}_q\, {\bf{v}}_q(f_q-f_{\bar{q}})}\nonumber\\
&-\sum_f\delta\omega_q2N_cq_{f_q}\int{d\Tilde{P}_q\,
\dfrac{{\bf{v}}_q}{\epsilon_{q}}(f_q-f_{\bar{q}})},
\end{align}
with $q_{f_{\bar{q}}}=-q_{f_{q}}$. We use the Boltzmann equation Eq.~(\ref{16}) to consider the $\delta f_k$ due to the external perturbation ${\bf E}$ in the weakly magnetized medium. Following the same formalism in Ref.~\cite{Feng:PRD962017} within the EQPM framework, we obtain
\begin{align}\label{34}
\delta f_k=&-q_{f_k}E v_{k\, x} \bigg(\frac{\partial f^0_k}{\partial\epsilon_k}\bigg)\frac{\tau_{R_k}}{(1+\tau_{R_k}^2\, \Omega^2_{c\, k})}\nonumber\\
&+q_{f_k}E v_{k\, y} \bigg(\frac{\partial f^0_k}{\partial\epsilon_k}\bigg)\frac{\Omega_{c\, k}\tau_{R_k}^2}{(1+\tau_{R_k}^2\, \Omega^2_{c\, k})}.
\end{align}    
Substituting Eq.~(\ref{34}) to Eq.~(\ref{33}) and employing Eq.~(\ref{32}) we obtain the electrical conductivity and Hall conductivity respectively as,
\begin{align}\label{35}
\sigma_e=&\frac{1}{3T}\sum_{k}g_k\tau_{R_k}q^2_{f_k}\int{d\Tilde{P}_k}\frac{\mid{\bf{\Tilde{p}}}_k\mid^2}{\omega^2_k}\frac{1}{(1+\tau_{R_k}^2\, \Omega^2_{c\, k})}\Big(-\frac{\partial f^0_k}{\partial\epsilon_k}\Big)\nonumber\\
&-\frac{1}{3T}\sum_{k}\delta\omega_kg_k\tau_{R_k} q^2_{f_k}\int{d\Tilde{P}_k}\frac{\mid{\bf{\Tilde{p}}}_k\mid^2}{\omega^2_k}\frac{1}{\epsilon_k}\frac{1}{(1+\tau_{R_k}^2\, \Omega^2_{c\, k})}\nonumber\\&\times\Big(-\frac{\partial f^0_k}{\partial\epsilon_k}\Big),
\end{align}
and 
\begin{align}\label{36}
\sigma_H=&\frac{1}{3T}\sum_{k}g_k\tau_{R_k}^2q^2_{f_k}\int{d\Tilde{P}_k}\frac{\mid{\bf{\Tilde{p}}}_k\mid^2}{\omega^2_k}\frac{1}{(1+\tau_{R_k}^2\, \Omega^2_{c\, k})}\nonumber\\&\times\Omega_{c\, k}\Big(-\frac{\partial f^0_k}{\partial\epsilon_k}\Big)\nonumber\\
&-\frac{1}{3T}\sum_{k=q,{\bar{q}}}\delta\omega_kg_k\tau_{R_k}^2 q^2_{f_k}\int{d\Tilde{P}_k}\frac{\mid{\bf{\Tilde{p}}}_k\mid^2}{\omega^2_k\, \epsilon_k}\frac{1}{(1+\tau_{R_k}^2\, \Omega^2_{c\, k})}\nonumber\\&\times\Omega_{c\, k}\Big(-\frac{\partial f^0_k}{\partial\epsilon_k}\Big).
\end{align}
Note that the heat current defined in Eq.~(\ref{28}) is analogous to the definition of electric charge current in Eq.~(\ref{32}). The thermal driving force (which is related to the temperature gradient) is the source of perturbation for the dissipative thermal transport process, whereas the electric field perturbs the medium for the charge transport.
In the presence of the magnetic field, in addition to the Ohmic current, the Lorenz force results in the Hall current perpendicular to the particle velocity and the magnetic field, and we define  $\sigma_H$ as the associated transport coefficient. The coefficient $\kappa_1$ is analogous to Hall conductivity $\sigma_H$ as the second term in Eq.~(\ref{28}) describes the heat current transverse to the thermal driving force and magnetic field. Similar to $\kappa_1$, $\sigma_H$ is in second order in thermal relaxation time and is subdominant in comparison with $\sigma_e$.
From Eq.~(\ref{36}) and Eq.~(\ref{32}), we observe that the second order current (Hall current) is proportional to $\mu$ and vanishes at $\mu=0$ limit. This observation is in line with that of Ref.~\cite{Satow:2014lia}. From Eq.~(\ref{30}), we see that $\kappa_1$ decreases with decrease in $\mu$ as the factor $\frac{q_{f_k}B}{\omega_k}$ depends on the charge of the particle species $k$.

Here, to compare the thermal dissipation with the electromagnetic responses in the medium, the Wiedemann-Franz law has been employed in two different directions by considering the thermal driving force and electric as the sources of perturbation for the dissipative thermal and electric charge transport processes, respectively. The presence of the electric field in the thermal transport or vice versa ($i.e.$, the thermal driving force in the electric charge transport) give more components of thermal and electric conductivities in the medium, that may define the tensorial structure of ${ L}$ as in the case of anisotropic crystals. The thermoelectric behaviour of the magnetized medium is beyond the scope of the current analysis.

\subsection{Weak magnetic field versus strong field limit }

In the presence of a strong magnetic field (along the $z-$ axis), the charged particle motion is constrained in the direction of the field via $1+1-$dimensional Landau level kinematics.   It has been shown that $1\rightarrow 2$ processes are kinematically possible in the dimensionally reduced medium in Ref.~\cite{Hattori:2017qih}. Thermal relaxation in the strongly magnetized hot QCD medium for the dominant $1\rightarrow 2$ processes has been estimated in Ref.~\cite{Kurian:2018qwb}. The lowest Landau level (LLL) approximation  is valid in the regime $T^{2}\ll \mid q_{f_k} B\mid$. However, the LLL approximation is questionable at high temperature regimes. In our previous works~\cite{Kurian:2018qwb,Kurian:2019fty}, we have estimated viscous coefficients, electrical conductivity, and thermal conductivity with full Landau level resummation in a more realistic regime $gT\ll \sqrt{\mid q_fB\mid}$. The thermal conductivity in the strongly magnetized medium takes the following form,
\begin{align}\label{37}
\kappa_0&= \sum_{l=0}^{\infty}\sum_{k}d_l\dfrac{\mid q_{f_k}B\mid}{2\pi}
\dfrac{N_c}{T^2}\int_{-\infty}^{\infty}{\dfrac{d\Tilde{p}_{z_k}}
{2\pi}\tau_{\text{eff}}\dfrac{(\omega^l_{k}-h^l_k)^2}
{\omega^{l\, 2}_{k}}}\nonumber\\
&\times\Tilde{p}_{z_k}^2
f^{l\,0}_k(1-f^{l\,0}_k) \nonumber\\
&-\sum_{l=0}^{\infty}\sum_{k}d_l\delta\omega_k\dfrac{\mid {q_{f_k}}B\mid}
{2\pi}\dfrac{N_c}{T^2}\int_{-\infty}^{\infty}{\dfrac{d\Tilde{p}_{z_k}}
{2\pi}\tau_{\text{eff}}\dfrac{h^l_k(\omega^l_{k}-h^l_k)}
{\omega^{l\, 2}_{k}}}\nonumber\\
&\times\dfrac{\Tilde{p}_{z_k}^2}{\sqrt{p_{z}^{2}+m_f^{2}+2l\mid q_{f_k}B\mid}}
f^{l\,0}_k(1-f^{l\,0}_k),
\end{align}
\begin{figure*}
 \centering
 \subfloat{\includegraphics[scale=0.4]{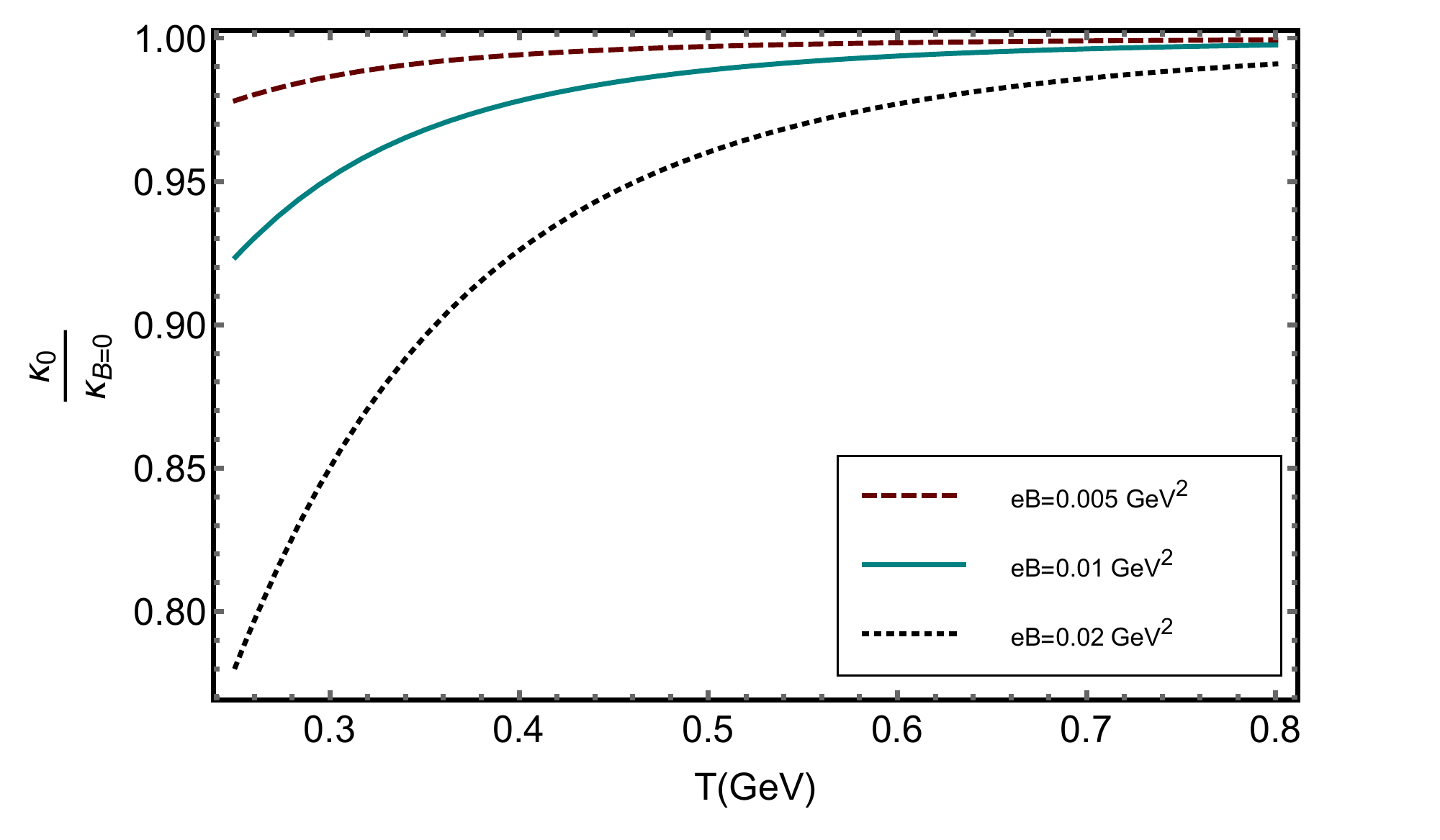}}
 \subfloat{\includegraphics[scale=0.4]{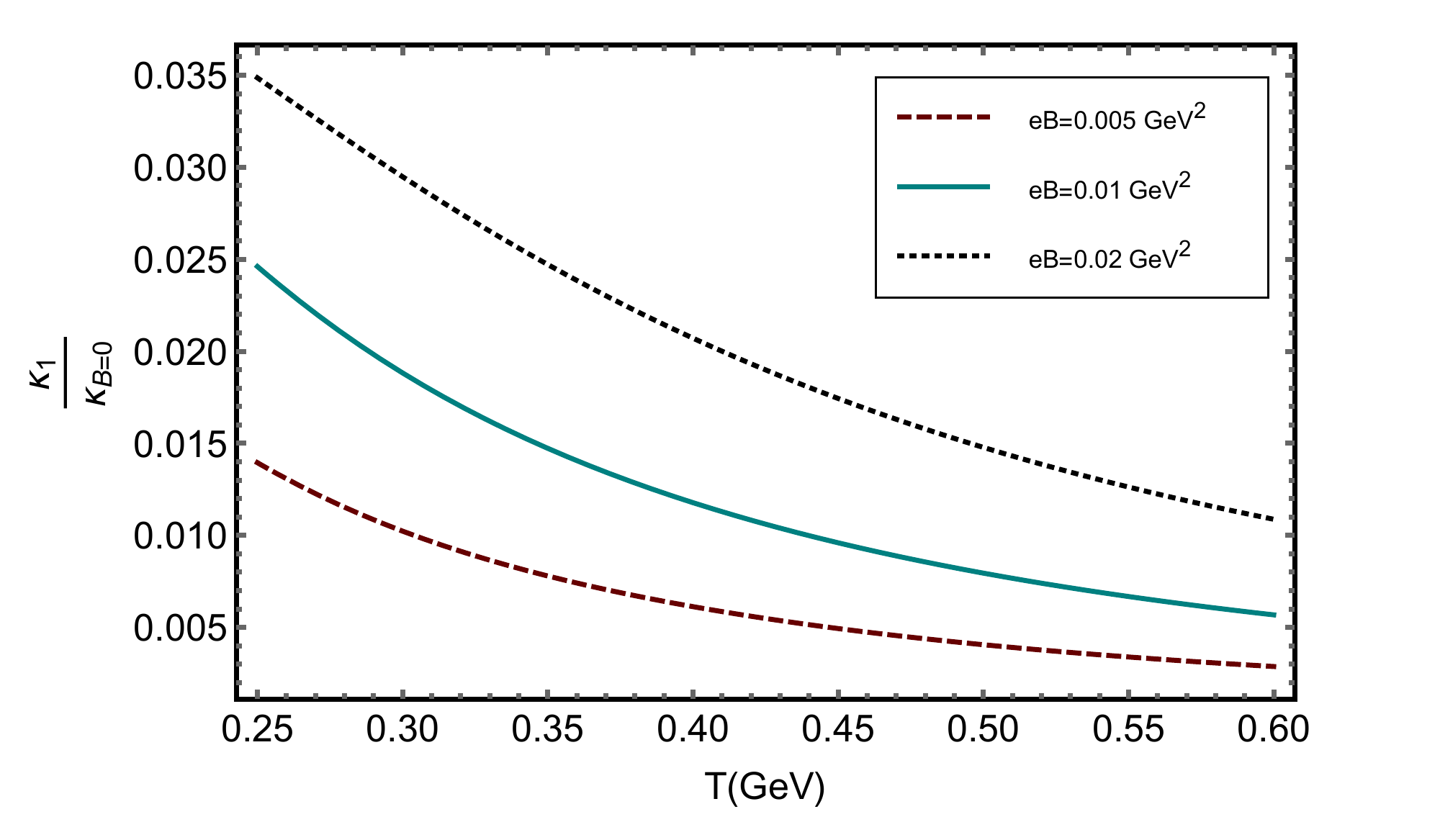}}
\caption{(Color online) The magnetic field dependence on the $\kappa_0$ (left panel) and $\kappa_1$ (right panel) at $\mu=200$ MeV.}
\label{f2}
\end{figure*}
where $d_l=(2-\delta_{l0})$ is the spin degeneracy factor of Landau levels $l$, $h^l_k$ is the enthalpy per particle in the presence of the strong magnetic field, and $\tau_{eff}$ is the relaxation time for the $1\rightarrow 2$ processes in the strong field limit. Here, $f^{l\,0}_k$ and $\omega_{l_k}$ denotes the quark/antiquark distribution function and single quasiparticle energy respectively and take the forms,
\begin{equation}\label{38}
f^{l\,0}_{k}=\dfrac{z_{k}\exp{\bigg[-\beta \Big(\sqrt{p_{z}^{2}+m_f^{2}+2l\mid q_{f_k}B\mid
}\mp\mu\Big)\bigg]}}{1+ z_{k}\exp{\bigg[-\beta \Big(\sqrt{p_{z}^{2}+m_f^{2}+2l\mid q_{f_k}B\mid
}\mp\mu\Big)\bigg]}},
\end{equation}
 and 
\begin{equation}\label{39}
\omega^l_{k}=
\sqrt{p_{z}^{2}+m_f^{2}+2l\mid q_{f_k}B\mid}+\delta\omega_k.
\end{equation} 
Note that in the strongly magnetized medium gluonic contribution to the thermal conductivity is negligible in comparison with quark and antiquark contribution. The heat current transverse to the magnetic field is negligible as the dominant contribution of temperature gradient in the dimensionally reduced medium is along the direction of the field, $i.e.$, $\kappa_1$ vanishes in the strong field limit. The same observation, $i.e.$, $\sigma_H\sim 0$, holds true for the charge transport in the strongly magnetized medium.

The magnetic field dependence on the temperature behaviour of the Lorenz number $L$ in a weakly magnetized medium is analyzed and compared the results with that in the case of the strong field regime (beyond the LLL approximation), in the next section.


\section{Results and discussions}

The mean field contributions to the thermal conductivity $\kappa_0$ and Hall type conductivity $\kappa_1$ associated with the thermal transport process in the weakly magnetized QGP is described in Eq.~(\ref{29}) and Eq.~(\ref{30}). The mean field effects that arise from the EQPM description of the medium to the transport coefficients are well investigated in Refs.~\cite{Mitra:2018akk,Bhadury:2019xdf}. At high temperature regimes, the system attains non-interacting ideal EoS (asymptotically free), $i.e.$,  $z_{g/q}\rightarrow 1$, at very high temperature. Hence, the mean field terms to $\kappa_0$ and $\kappa_1$ are negligible in the high temperature regimes.

The effect of the magnetic field on $\kappa_0$ and $\kappa_1$ are shown in Fig.\ref{f2}. Note that the current analysis is on the weakly magnetized medium with temperature as the dominant energy scale in the system as compared to the magnetic field. The magnetic field dependence on the thermal conductivity $\kappa_0$ is entering through the cyclotron frequency $\mid \Omega_{c\, k}\mid$ as described in Eq.~(\ref{29}). The temperature behaviour of the ratio of $\kappa_0$ to the thermal conductivity in the absence of magnetic field $\kappa_{B=0}$ is plotted at $\mid eB\mid=0.005$ GeV$^2$, $0.01$ GeV$^2$ and $0.02$ GeV$^2$ at a finite chemical potential $\mu=200$ MeV (left panel). It is seen that the coefficient $\kappa_0$ decreases with an increase in the strength of the magnetic field. This is due to the factor $\frac{1}{1+\tau_{R_k}^2\, \Omega^2_{c\, k}}$ that originates from the Lorenz force term in the Boltzmann equation in the presence of a weak magnetic field. The effect of the magnetic field is more visible in the lower temperature regimes. The results reduced to that of Ref.\cite{Chakrabarty:1986xx} in the case of vanishing magnetic field at the ultra-relativistic limit ($z_k\rightarrow 1$). Due to the Lorenz force, the initial motion of the particle gets deflected in a weakly magnetized medium. This, in turn, leads to another component of thermal transport coefficient $\kappa_1$ in the direction perpendicular to the magnetic field and thermal driving force at a finite chemical potential. The temperature dependence of $\kappa_1$ at $\mu=200$ MeV is depicted in the right panel of Fig.\ref{f2}. The coefficient $\kappa_1$ increases with the strength of the magnetic field as it is proportional to the factor $\frac{\Omega_{c\, k}}{1+\tau_{R_k}^2\, \Omega^2_{c\, k}}$. The ratio approaches zero asymptotically, and this implies that the effect of the  magnetic field in the thermal transport in a weakly magnetized medium is negligible at a very high temperature. It is important to emphasize that the coefficient $\kappa_1$ vanishes in the dimensionally reduced system in the presence of a strong 
magnetic field. 
\begin{figure}[h]
\hspace{-1cm}
  \subfloat{\includegraphics[scale=0.4]{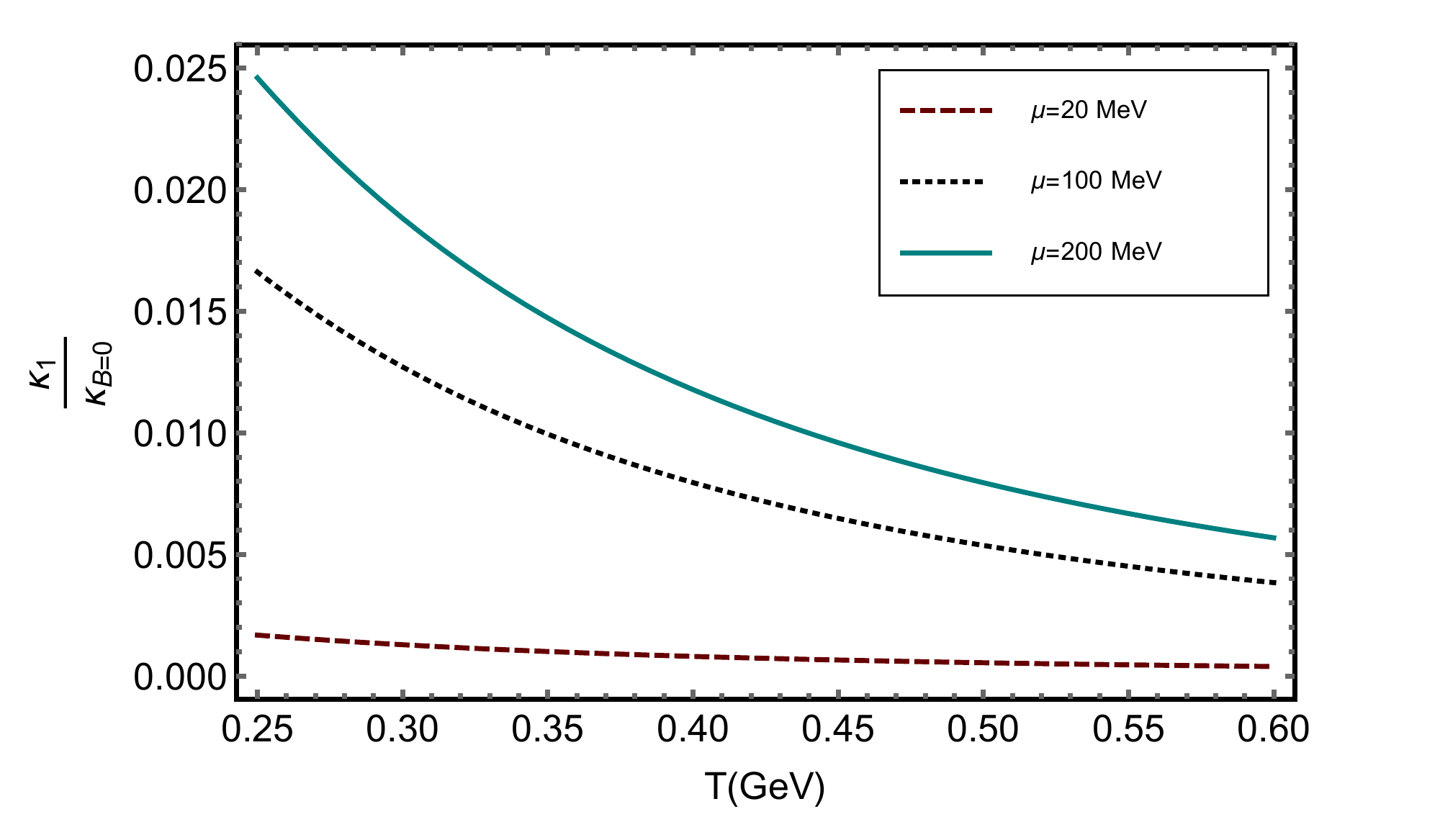}}
\caption{(Color online) Dependence of chemical potential on the temperature behaviour of $\kappa_1$ at $\mid eB\mid=0.01$ GeV$^2$.}
\label{f3}
\end{figure}
\begin{figure*}
 \centering
 \subfloat{\includegraphics[scale=0.4]{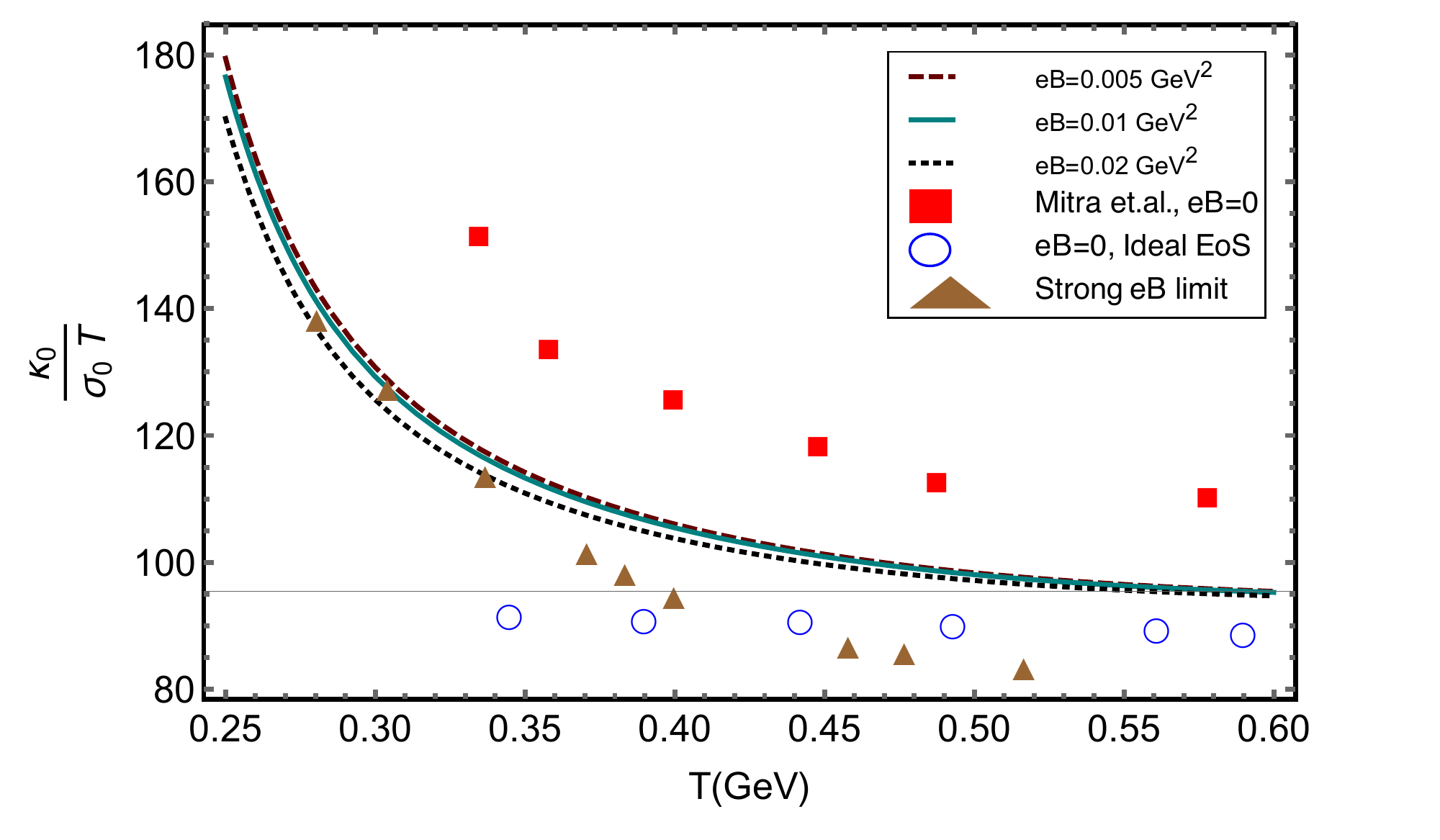}}
 \subfloat{\includegraphics[scale=0.4]{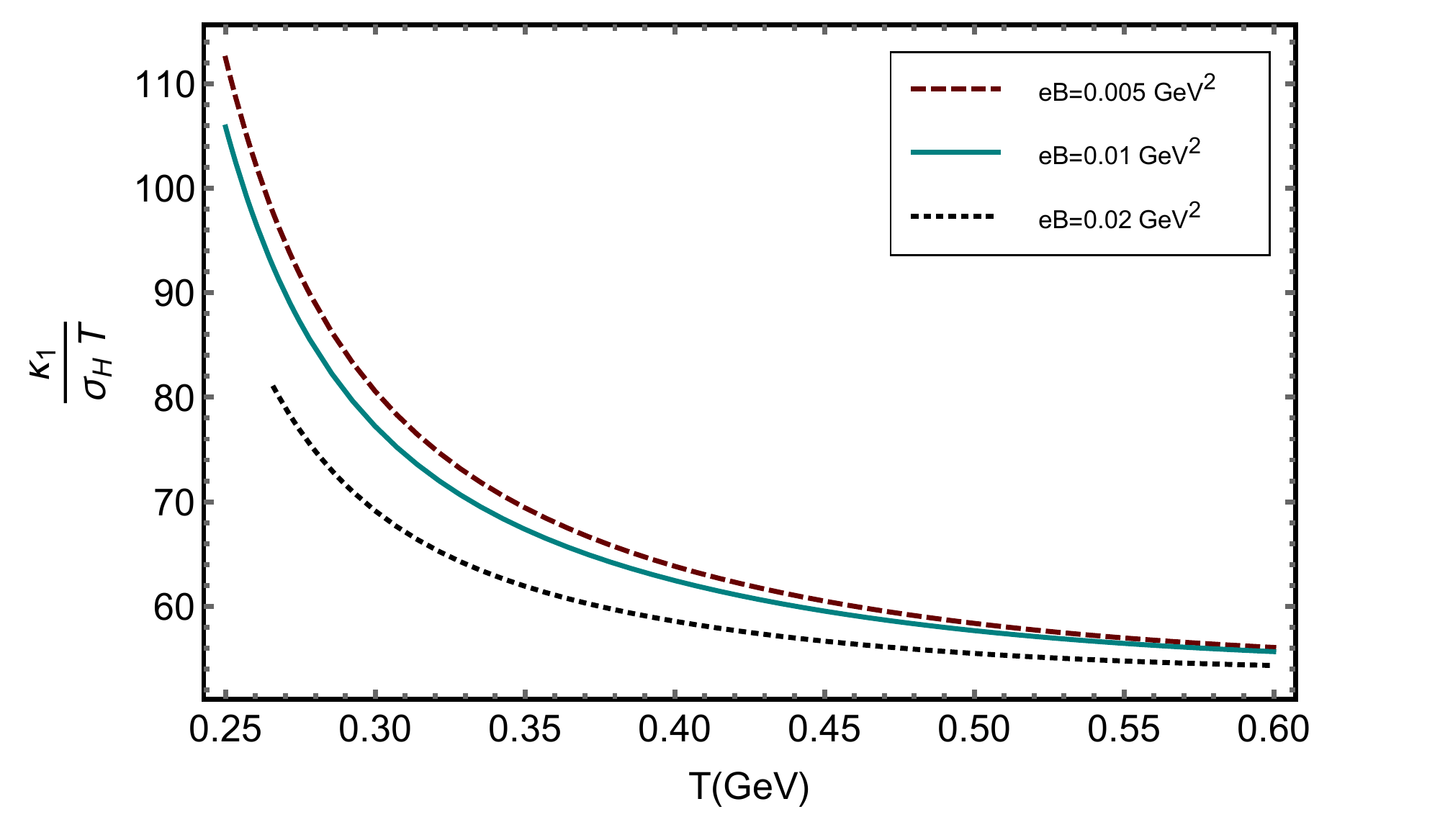}}
\caption{(Color online) 
Lorenz number within the EQPM as a function of temperature in a weakly magnetized medium at $\mid eB\mid=0.01$ GeV$^2$ and $\mu=200$ MeV. }
\label{f4}
\end{figure*}

In Fig.\ref{f3}, the dependence of finite quark chemical potential on temperature behaviour of $\kappa_1$ is plotted for a fixed magnetic field, say $\mid eB\mid=0.01$ GeV$^2$. We observe that the ratio of Hall type thermal transport coefficient to $\kappa_{B=0}$ decreases with a decrease in $\mu$. This can be understood from Eq.~(\ref{30}) as $\Omega_{c\, k}=\frac{q_{f_k}B}{\omega_k}$ depends on the charge of the species and we have, $q_{f_q}=-q_{f_{\bar{q}}}$. Note that the cyclotron frequency $\frac{\mid q_{f_k}B\mid}{\omega_k}$ remains intact for both species.
A similar observation holds true for the Hall conductivity associated with the electric charge transport in a weakly magnetized medium~\cite{Satow:2014lia}.

The relative behaviour of thermal transport and electric charge transport in the weakly magnetized medium is quantified in terms of Lorenz number $L$. The temperature behaviour of the Lorenz number for the weakly magnetized QGP is plotted in the directions transverse to the direction of the magnetic field (say, $x-$axis and $y-$axis, in the current analysis) in Fig.\ref{f4}. For the temperature range from $T=0.25$ GeV to  $T=0.5$ GeV, $L\equiv\frac{\kappa_0}{\sigma_eT}$ varies from $160$ to $95$ for the weakly magnetized medium. At high temperature regime $T>0.45$ GeV, the magnitude of $L$ saturates closer to $95$, which is the Stefan-Boltzmann limit of the QGP (ultra-relativistic limit). It is observed that the Wiedemann-Franz law is violated in the weakly magnetized medium at the lower temperature regime. This observations perhaps indicates towards much more complex behaviour of the QGP medium as a strongly interacting quantum fluid near the transition temperature. Noticeably, deviation of the Lorenz number has been studied in holographic anisotropic models, which are dual to $N = 4$ spatially anisotropic Super Yang-Mills (SYM) theory at finite $\mu$~\cite{Ge:2014aza} and also in a color string percolation scenario~\cite{Sahoo:2019xjq}. Recently, the authors of Ref.\cite{Rath:2019vvi} have realized the violation of the Wiedemann-Franz law in the presence of a strong magnetic field within LLL approximation. In the current analysis, we have extended the analysis of temperature dependence of Lorenz number from the weak magnetic field limit to the strong field limit while considering the higher Landau level contributions to the longitudinal electrical and thermal conductivities within the EQPM. We have compared the results with that of Ref.~\cite{Mitra:2017sjo} where the validity of the Wiedemann-Franz law is studied for the medium with non-ideal EoS at vanishing magnetic field without incorporating the mean field effects. It is seen that both the magnetic field and EoS have a significant impact on the Lorenz number both in the weak and strong magnetic fields limit. We have observed similar temperature behaviour for the Lorenz number $L\equiv\frac{\kappa_1}{\sigma_HT}$ along the transverse direction to the magnetic field and thermal driving force (and electric field), say along the $y-$axis. At a very high temperature, the number saturates closer to $55$ within the EQPM description.

\section{Conclusion and Outlook}

In this article, we have studied the thermal transport of the medium in the presence of a weak magnetic field. We have incorporated the realistic EoS effects through the EQPM modeling of the equilibrium distribution functions by considering the system as a grand canonical ensemble of quarks, antiquarks, and quarks with a finite $\mu$. The effective covariant transport equation is employed to describe the evolution of the medium. The magnetic field is considered to be weak and is subdominant to the temperature energy scale in the system. We have studied the effects of mean field corrections to the transport coefficients associated with the thermal transport in the weakly magnetized medium at finite baryon chemical potential.  The presence of the magnetic field in the QGP medium leads to Hall type conductivity, which is transverse to both the magnetic field and thermal driving force, associated with the thermal transport process. We have studied the dependence of the magnetic field and quark chemical potential in the temperature behaviour of the thermal transport coefficients in the medium.  We found that the effects of the magnetic field and chemical potential are pronounced in the temperature regime near to the transition temperature. 

Further, we have investigated the relative significance of thermal and electric transport processes in the weakly magnetized medium in terms of the Wiedemann-Franz law. It is observed that the Wiedemann-Franz law is violated in the medium, especially in the lower temperature regimes. We have extended the analysis to the strong magnetic field limit while incorporating the higher Landau level effects. The temperature dependence of the Lorenz number has estimated for strong and weak magnetic fields. We have compared the results with other parallel works. It is observed that both the magnetic field and hot QCD medium interactions have key roles in the relative behaviour of thermal and electric charge transport of the medium, especially in the temperature regimes not very far from the transition temperature. 

However, heat transport induces electric transport and vice
versa. These aspects can be studied within the scope of the thermoelectric effect (Seebeck effect). The thermoelectric behavior of the magnetized QGP and the associated magneto-Seebeck coefficient and Nernst coefficient of the hot QCD medium within relaxation time approximation and BGK collision kernels are worth investigating. The present analysis of thermal transport in the magnetized medium to be the first step in this direction. 
The investigation of all components of shear and bulk viscous coefficients in a weakly magnetized medium within the effective kinetic theory
 and the effects of inhomogeneity of the electromagnetic fields to the transport coefficients are other interesting directions to work in the near future. 

\section*{acknowledgments}
The author acknowledges Vinod Chandra for immense encouragement, helpful discussions, and suggestions. The author further record a deep sense of gratitude to the people of India for their generous support for the research in fundamental sciences.



{}

\end{document}